\documentclass[10pt,conference]{IEEEtran}

\usepackage{epsfig}
\usepackage{amsmath}    
\usepackage{graphicx}
\usepackage{amssymb}

\begin{document}

\title{Capacity of the Bosonic Wiretap Channel and the Entropy Photon-Number Inequality}

\author{
\authorblockN{Saikat Guha}
\authorblockA{Research Laboratory of Electronics \\
MIT, Cambridge, MA 02139 \\
saikat@MIT.edu}
\and
\authorblockN{Jeffrey H. Shapiro}
\authorblockA{Research Laboratory of Electronics \\
MIT, Cambridge, MA 02139 \\
jhs@MIT.edu}
\and
\authorblockN{Baris I. Erkmen}
\authorblockA{Research Laboratory of Electronics \\
MIT, Cambridge, MA 02139 \\
erkmen@MIT.edu}
}

\maketitle

\begin{abstract}
Determining the ultimate classical information carrying capacity of electromagnetic waves requires quantum-mechanical analysis to properly account for the bosonic nature of these waves.  Recent work has established capacity theorems for bosonic single-user and broadcast channels, under the presumption of two minimum output entropy conjectures. Despite considerable accumulated evidence that supports the validity of these conjectures, they have yet to be proven. In this paper, it is shown that the second conjecture suffices to prove the classical capacity of the bosonic wiretap channel, which in turn would also prove the quantum capacity of the lossy bosonic channel.  The preceding minimum output entropy conjectures are then shown to be simple consequences of an Entropy Photon-Number Inequality (EPnI), which is a conjectured quantum-mechanical analog of the Entropy Power Inequality (EPI) form classical information theory. 
\end{abstract}

\section{Motivation and History}

The performance of communication systems that rely on electromagnetic wave propagation are ultimately limited by noise of quantum-mechanical origin.  Moreover, high-sensitivity photodetection systems have long been close to this noise limit. Hence determining the ultimate capacities of lasercom channels is of immediate relevance.  The most famous channel capacity formula is Shannon's result for the classical additive white Gaussian noise channel.  For a complex-valued channel model in which we transmit $a$ and receive $c = \sqrt{\eta}\,a + \sqrt{1-\eta}\,b$, where $0< \eta < 1$ is the channel's transmissivity and $b$ is a zero-mean, isotropic, complex-valued Gaussian random variable that is independent of $a$, Shannon's capacity is
\begin{equation}
C_{\rm classical} = \ln[1 + \eta\bar{N}/(1-\eta)N]\,\mbox{ nats/use},
\label{Shannon}
\end{equation}
with $E(|a|^2) \le \bar{N}$ and $E(|b|^2) = N$.  
In the quantum version of this channel model, we control the state of an electromagnetic mode with photon annihilation operator $\hat{a}$ at the transmitter, and receive another mode with photon annihilation operator $\hat{c} = \sqrt{\eta}\,\hat{a} + \sqrt{1-\eta}\,\hat{b}$, where $\hat{b}$ is the annihilation operator of a noise mode that is in a zero-mean, isotropic, complex-valued Gaussian state.  For lasercom, if quantum measurements corresponding to ideal optical homodyne or heterodyne detection are employed at the receiver, this quantum channel reduces to a real-valued (homodyne) or complex-valued (heterodyne) additive Gaussian noise channel, from which the following capacity formulas (in nats/use) follow:
\begin{eqnarray}
C_{\rm homodyne} &=& 2^{-1}\ln[1+ 4\eta\bar{N}/(2(1-\eta)N + 1)]
\label{homodyne}\\[.06in]
C_{\rm heterodyne} &=& \ln[1+ \eta\bar{N}/((1-\eta)N+1)].
\label{heterodyne}
\end{eqnarray}
The +1 terms in the noise denominators are quantum contributions, so that even when the noise mode $\hat{b}$ is unexcited these capacities remain finite, unlike the situation in 
Eq.~(\ref{Shannon}).  

The classical capacity of the pure-loss bosonic channel---in which the $\hat{b}$ mode is unexcited ($N = 0$)---was shown in \cite{ultcap} to be $C_{\rm pure-loss} = g(\eta\bar{N})$ nats/use, where $g(x) \equiv (x+1)\ln(x+1) - x\ln(x)$ is the Shannon entropy of the Bose-Einstein probability distribution with mean $x$.  This capacity exceeds the $N = 0$ versions of Eqs.~(\ref{homodyne}) and (\ref{heterodyne}), as well as the best known bound on the capacity of ideal optical direct detection.  The ultimate capacity of the thermal-noise ($N > 0$) version of this channel is bounded below as follows, $C_{\rm thermal} \ge g(\eta\bar{N} + (1-\eta)N) - g((1-\eta)N)$, and this bound was shown to be the capacity if the thermal channel obeyed a certain minimum output entropy conjecture \cite{thermal}. This conjecture states that the von Neumann entropy at the output of the thermal channel is minimized when the $\hat{a}$ mode is in its vacuum state.   Considerable evidence in support of this conjecture has been accumulated \cite{gglms}, but it has yet to be proven. Nevertheless, the preceding lower bound already exceeds Eqs.~(\ref{homodyne}) and (\ref{heterodyne}) as well as the best known bounds on the capacity of direct detection.  

More recently, a capacity analysis of the bosonic broadcast channel led to an inner bound on the capacity region, which was shown to be the capacity region under the presumption of a second minimum output entropy conjecture  \cite{broadcast}. Both conjectures have been proven if the input states are restricted to be Gaussian, and they have been shown to be equivalent under this input-state restriction.  In this paper, we show that the second conjecture will establish the privacy capacity of the lossy bosonic channel, as well as its ultimate quantum information carrying capacity.

The Entropy Power Inequality (EPI) from classical information theory is widely used in coding theorem converse proofs for Gaussian channels.  By analogy with the EPI, we conjecture its quantum version, viz., the Entropy Photon-number Inequality (EPnI).  In this paper we show that the two minimum output entropy conjectures cited above are simple corollaries of the EPnI.  Hence, proving the EPnI would immediately establish key results for the capacities of bosonic communication channels.

\section{Quantum Wiretap Channel}

The term ``wiretap channel" was coined by Wyner \cite{wyner}  to describe a communication system, in which Alice wishes to communicate classical information to Bob, over a point-to-point discrete memoryless channel that is subjected to a wiretap by an eavesdropper Eve. Alice's goal is to reliably and securely communicate classical data to Bob, in such a way that Eve gets no information whatsoever  about the message. Wyner used the conditional entropy rate of the signal received by Eve, given Alice's transmitted message, to measure the secrecy level guaranteed by the system. He gave a single letter characterization of the rate-equivocation region under a limiting assumption, that the signal received by Eve is a degraded version of the one received by Bob. Csisz\'{a}r and K\"{o}rner later generalized Wyner's results to the case in which the signal received by Eve is not a degraded version of the one received by Bob \cite{csiszar}. These classical-channel results were later extended  by Devetak \cite{devetak} to encompass classical transmission over a quantum wiretap channel.

A quantum channel ${\cal N}_{A\mbox{-}B}$ from Alice to Bob is a trace-preserving completely positive map that transforms Alice's single-use density operator ${{\hat{\rho}}}^A$ to Bob's, ${{\hat{\rho}}}^B = {\cal N}_{A\mbox{-}B}({\hat{\rho}}^A)$. The quantum wiretap channel ${\cal N}_{A\mbox{-}BE}$ is a quantum channel from Alice to an intended receiver Bob and an eavesdropper Eve . The quantum channel from Alice to Bob is obtained by tracing out $E$ from the channel map, i.e., ${\cal N}_{A\mbox{-}B} \equiv {\rm Tr}_E\left({\cal N}_{A\mbox{-}BE}\right)$, and similarly for ${\cal N}_{A\mbox{-}E}$. A quantum wiretap channel is degraded if there exists a degrading channel ${\cal N}^{\rm {deg}}_{B\mbox{-}E}$ such that ${\cal N}_{A\mbox{-}E} = {\cal N}^{\rm {deg}}_{B\mbox{-}E} \circ {\cal N}_{A\mbox{-}B}.$

The wiretap channel describes a physical scenario in which for each successive $n$ uses of ${\cal N}_{A\mbox{-}BE}$ Alice communicates a randomly generated classical message $m \in W$ to Bob, where $m$ is a classical index that is uniformly distributed over the set, $W$, of $2^{nR}$ possibilities. To encode and transmit $m$, Alice generates an instantiation $k \in K$ of a discrete random variable, and then prepares $n$-channel-use states that after transmission through the channel, result in bipartite conditional density operators $\{{\hat \rho}_{m,k}^{B^nE^n}\}$.   A $(2^{nR}, n, \epsilon)$ code for this channel consists of an encoder,
$x^n: (W,K) \rightarrow {\cal A}^n$,
and a positive operator-valued measure (POVM) $\{\Lambda_{m}^{B^n}\}$ on ${\cal{B}}^n$ such that the following conditions are satisfied for every $m \in W$.\footnote[1]{${\cal A}^n$, ${\cal B}^n$, and ${\cal E}^n$ are the $n$-channel-use alphabets of Alice, Bob and Eve.}
\begin{enumerate}
\item Bob's probability of decoding error is at most $\epsilon$, i.e., 
\begin{equation}
{\rm Tr}\!\left(\hat{\rho}_{m,k}^{B^n}\Lambda_{m}^{B^n}\right) > 1-\epsilon, \quad \forall k, \quad {\text{and}}
\label{eq:dec_condition1}
\end{equation}

\item For any POVM $\{\Lambda_{m}^{E^n}\}$ on ${\cal E}^n$, no more than $\epsilon$ bits of information is revealed about the secret message $m$. Using $j \equiv (m,k)$, this condition can be expressed, in terms of the Holevo information \cite{holevo}, as follows, 
\begin{equation}
\chi\!\left(p_j,{\cal N}_{A-E}^{\otimes n}({\rho_j^{A^n}})\right) \le \epsilon.
\label{eq:dec_condition2}
\end{equation}
Here,
$
\chi({p_j, {  {\hat{\sigma}}}_j}) \equiv S(\sum_j{p_j{  {\hat{\sigma}}}_j}) - \sum_jp_jS({  {\hat{\sigma}}}_j),
$ is the Holevo information, where
$\{p_j\}$ is a probability distribution associated with the density operators ${  {\hat{\sigma}}}_j$, and $S({\hat{\rho}}) \equiv -{\rm {Tr}}({\hat{\rho}}\log{\hat{\rho}})$ is the von Neumann entropy of the density operator ${\hat{\rho}}$.\footnote[2]{A density operator is Hermitian, with eigenvalues that form a probability distribution.  Thus, the von Neumann entropy of a density operator $\hat{\rho}$ is the Shannon entropy of its eigenvalues.} 

\end{enumerate}

Because Holevo information may not be additive, the classical privacy capacity $C_p$ of the quantum wiretap channel must be computed by maximizing over successive uses of the channel, i.e., for $n$ being the number of uses of the channel,
\begin{eqnarray}
&&C_p({\cal N}_{A\mbox{-}BE}) \nonumber \\
&&= \sup_n\max_{p_T(i)p_{A|T}(j|i)}\left[\chi(p_T(i),\mbox{$\sum_j$}p_{A|T}(j|i){\hat \rho}_j^{B^n})/n \right. \nonumber \\
&&-\left. \chi(p_T(i),\mbox{$\sum_j$}p_{A|T}(j|i){\hat \rho}_j^{E^n})/n\right]. \label{eq:privCap} 
\end{eqnarray}
The probabilities $\{p_i\}$ form a distribution over an auxiliary classical alphabet ${\cal T}$, of size $|{\cal T}|$. The ultimate privacy capacity is computed by maximizing the expression specified in \eqref{eq:privCap} over $\{p_T(i)\}$, $\{p_{A|T}(j|i)\}$, $\{{\hat \rho}_j^{A^n}\}$, and $n$, subject to a cardinality constraint on $|{\cal T}|$. For a degraded wiretap channel, the auxiliary random variable is unnecessary, and Eq.~\eqref{eq:privCap} reduces to
\begin{equation}
C_p({\cal{N}}_{A\mbox{-}BE}) = \sup_n\max_{p_{A}(j)}[\chi(p_A(j),{\hat \rho}_j^{B^n})/n - \chi(p_A(j),{\hat \rho}_j^{E^n})/n].
\label{eq:privCap_degraded}
\end{equation}

\section{Noiseless Bosonic Wiretap Channel}

The noiseless bosonic wiretap channel consists of a collection of spatial and temporal bosonic modes at the transmitter that interact with a minimal-quantum-noise environment and split into two sets of spatio-temporal modes en route to two independent receivers, one being the intended receiver and the other being the eavesdropper. The multi-mode bosonic wiretap channel is given by $\bigotimes_s{{\cal N}_{A_s\mbox{-}B_sE_s}}$, where ${{\cal N}_{A_s\mbox{-}B_sE_s}}$ is the wiretap-channel map for the $s$th mode, which can be obtained from the Heisenberg evolutions
\begin{eqnarray}
{\hat b}_s &=& {\sqrt {\eta_s}}\,{\hat a}_s + {\sqrt {1-\eta_s}}\,{\hat f}_s, {\quad} \label{eq:WS-modebk} \\
{\hat e}_s &=& {\sqrt {1-\eta_s}}\,{\hat a}_s - {\sqrt {\eta_s}}\,{\hat f}_s, \label{eq:WS-modeek}
\end{eqnarray}
where the $\{{\hat a}_s\}$ are Alice's modal annihilation operators, and $\{{\hat b}_s\}$, $\{{\hat e}_s\}$ are the corresponding modal annihilation operators for Bob and Eve, respectively. The modal transmissivities $\{\eta_s\}$ satisfy $0 \le \eta_s \le 1$, and the environment modes $\{{\hat f}_s\}$ are in their vacuum states. We will limit our treatment here to the single-mode bosonic wiretap channel, as the privacy capacity of the multi-mode channel can in principle be obtained by summing up capacities of all spatio-temporal modes and maximizing the sum capacity subject to an overall input-power budget using Lagrange multipliers, cf.~\cite{thermal}, where this was done for  the multi-mode single-user lossy bosonic channel. 

\begin{figure}
\begin{center}
\includegraphics[width=6cm]{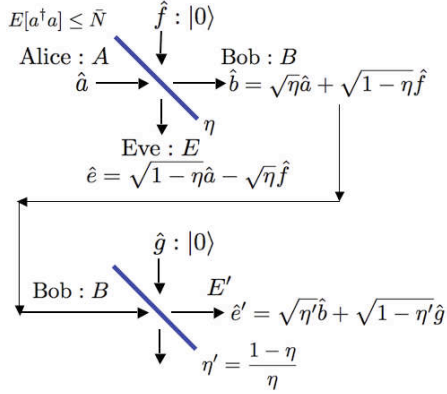}
\end{center}
\caption{Schematic diagram of the single-mode bosonic wiretap channel. The transmitter Alice ($A$) encodes her messages to Bob ($B$) in a classical index $j$, and over $n$ successive uses of the channel, thus preparing a bipartite state ${\hat \rho}_j^{B^nE^n}$ where $E^n$ represents $n$ channel uses of an eavesdropper Eve ($E$). For $\eta > 1/2$, this channel is degraded, as Eve's state can be recreated by passing Bob's state through a beamsplitter of transmissivity $(1-\eta)/\eta$.}
\label{fig:boswiretap}
\end{figure}

\noindent {\bf Theorem} --- Assuming the truth of minimum output entropy conjecture 2 (see Sec.~V), the ultimate privacy capacity of the single-mode noiseless bosonic wiretap channel (see Fig.~\ref{fig:boswiretap}) with mean input photon-number constraint $\langle{\hat a}^\dagger{\hat a}\rangle \le {\bar N}$ is 
\begin{equation}
C_p({\cal N}_{A\mbox{-}BE}) = g(\eta{\bar N}) - g((1-\eta){\bar N})\,\mbox{ nats/use},
\label{eq:Cp_ult}
\end{equation}
for $\eta > 1/2$ and $C_p = 0$ for $\eta \le 1/2$. This capacity is additive and achievable with single-channel-use coherent-state encoding with a zero-mean isotropic Gaussian prior distribution $p_{A}(\alpha) = \exp(-{|\alpha|^2}/{{\bar N}})/\pi \bar{N}$.\\

\noindent {\bf Proof} --- Devetak's result for the privacy capacity of the degraded quantum wiretap channel in Eq.~\eqref{eq:privCap_degraded} requires finite-dimensional Hilbert spaces. Nevertheless, we will use this result for the bosonic wiretap channel, which has an infinite-dimensional state space, by extending it to infinite-dimensional state spaces through a limiting argument.\footnote[3]{When $|{\cal T}|$ and $|{\cal A}|$ are finite and we are using coherent states in Eq.~\eqref{eq:privCap_degraded}, there will be a finite number of possible transmitted states, leading to a finite number of possible states received by Bob and Eve.  Suppose we limit the auxiliary-input alphabet ($T$)---and hence the input ($A$) and the output alphabets ($B$ and $E$)---to truncated coherent states within the finite-dimensional Hilbert space spanned by the Fock states $\{\,|m\rangle : 0 \le m \le M\,\}$, where $M \gg {\bar N}$. Applying Devetak's theorem to the Hilbert space spanned by these truncated coherent states then gives us a lower bound on the privacy capacity of the bosonic wiretap channel when the entire, infinite-dimensional Hilbert space is employed.  By taking $M$ sufficiently large, while maintaining the cardinality condition for ${\cal{T}}$, the rate-region expressions given by Devetak's theorem will converge to Eq.~\eqref{eq:Cp_ult}.}  Furthermore, it was recently shown that the privacy capacity of a degraded wiretap channel is additive, and equal to the single-letter quantum capacity of the channel from Alice to Bob \cite{graeme}, i.e.,
\begin{equation}
C_p({\cal N}_{A\mbox{-}BE}) = C_p^{(1)}({\cal N}_{A\mbox{-}BE}) = Q^{(1)}({\cal N}_{A\mbox{-}B}),
\label{eq:graeme}
\end{equation}
where the superscript ${(1)}$ denotes single-letter capacity.  
It is straightforward to show that if $\eta > 1/2$, the bosonic wiretap channel is a degraded channel, in which Bob's is the less-noisy receiver and Eve's is the more-noisy receiver. The degraded nature of the bosonic wiretap channel has been depicted in Fig.~\ref{fig:boswiretap}, where the quantum states ${\hat \rho}^{E^\prime}$ of the constructed system $E^\prime$ are identical to the quantum states ${\hat \rho}^{E}$ for a given input quantum state ${\hat \rho}^A$. Using Eq.~\eqref{eq:graeme} for the bosonic wiretap channel, we have
\begin{eqnarray} 
\lefteqn{C_p({\cal N}_{A\mbox{-}BE}) = \max_{\langle \hat{a}^\dagger \hat{a}\rangle \le {\bar N}} \left[S\left({\hat \rho}^B\right) - S\left({\hat \rho}^E\right)\right]}  \nonumber \\[.06in]
&=& \max_{\langle
\hat{a}^\dagger\hat{a}\rangle \le \bar{N}} [S({\hat \rho}^B) - S({\hat \rho}^{E^\prime})] \nonumber \\[.06in]
&=& \max_{0 \le K \le g(\eta{\bar N})}
\{\mbox{$\max_{
\langle \hat{a}^\dagger\hat{a}\rangle \le \bar{N}, 
S({\hat \rho}^B) = K}$}
[S({\hat \rho}^B) - S({\hat \rho}^{E^\prime})]\} 
\nonumber \\[.06in]
&=& \max_{0 \le K \le g(\eta{\bar N})}\{K - \mbox{$\min_{
\langle \hat{a}^\dagger\hat{a}\rangle \le \bar{N}, S({\hat \rho}^B) = K}$} 
[S({\hat \rho}^{E^\prime})]\} \nonumber \\[.06in]
&=&\max_{0 \le K \le g(\eta{\bar N})}\{K - g[(1-\eta)g^{-1}(K)/\eta]\} \nonumber \\[.06in]
&=&g(\eta{\bar N})-g((1-\eta){\bar N})\,\mbox{ nats/use} \nonumber \\[.06in]
&=&Q^{(1)}({\cal N}_{A\mbox{-}B}).
\end{eqnarray}
The first equality above follows from Lemma 3 of \cite{graeme}. The second equality follows from ${\cal N}_{A\mbox{-}BE}$ being a degraded channel. The restriction to $0 \le K \le g(\eta\bar{N})$  in the third equality is permissible because $\max_{\langle \hat{a}^\dagger \hat{a}\rangle \le \bar N}S({\hat \rho}^B) = g({\eta}{\bar N})$. The fifth equality follows\footnote[4]{Here, $g^{-1}(S)$ is the inverse of the function $g(N)$. Because $g(N)$ for $N \ge 0$ is a non-negative, monotonically increasing, concave function of $N$, it has an inverse, $g^{-1}(S)$ for $S \ge 0$, that is non-negative, monotonically increasing, and convex.} from minimum output entropy conjecture 2 (see Sec.~V). The $\hat{\rho}^B$ that achieves this equality is a thermal state, which is realized when Alice employs coherent-state encoding with a zero-mean isotropic Gaussian prior distribution $p_A(\alpha) = (1/\pi K)\exp(-|\alpha|^2/K)$.  The sixth equality now follows from $g(x) - g(c x)$ being a monotonically increasing function of $x \ge 0$, for $c$ a constant satisfying $0 \le c < 1$,
and the equality to the single-letter quantum capacity follows from Eq.~\eqref{eq:graeme}. Note that the privacy capacity of this channel is zero when $\eta \le 1/2$. It is straightforward to show that in the limit of high input photon number $\bar N$, 
\begin{equation}
C_p({\cal N}_{A\mbox{-}BE}) = Q^{(1)}({\cal N}_{A\mbox{-}B}) = \max\left\{0, \ln(\eta) - \ln(1-\eta) \right\}, \nonumber
\end{equation}
a result that Wolf et. al. \cite{wolf} independently derived by a different approach without use of an unproven output entropy conjecture.  

\section{The Entropy Photon-Number Inequality (EPnI)}

\subsection{The Entropy Power Inequality}
Let $\bf X$ and $\bf Y$ be statistically independent, $n$-dimensional, real-valued random vectors that possess differential (Shannon) entropies $h({\bf X})$ and $h({\bf Y})$ respectively. Because a real-valued, zero-mean Gaussian random variable $U$ has differential entropy given by $h(U) = \ln(2\pi e \langle U^2\rangle)$, where the mean-squared value, $\langle U^2\rangle$, is considered to be the \em power\/\rm\ of $U$, the entropy powers of ${\bf X}$ and ${\bf Y}$ are taken to be
\begin{equation}
P({\bf X}) \equiv \frac{e^{h({\bf X})/n}}{2\pi e}\quad\mbox{and}\quad
P({\bf Y}) \equiv \frac{e^{h({\bf Y})/n}}{2\pi e}.
\end{equation}
In this way, an $n$-dimensional, real-valued, random vector $\tilde{\bf X}$ comprised of independent, identically distributed (i.i.d.), real-valued, zero-mean, variance-$P({\bf X})$, Gaussian random variables has differential entropy $h(\tilde{\bf X}) = h({\bf X})$.  We can similarly define an i.i.d. Gaussian random vector $\tilde{\bf Y}$ with differential entropy $h(\tilde{\bf Y}) = h({\bf Y})$. We define a new random vector by the convex combination 
\begin{equation}
{\bf Z} \equiv \sqrt{\eta}\,{\bf X} + \sqrt{1-\eta}\,{\bf Y}, 
\end{equation}
where $0\le \eta \le 1$.  This random vector has differential entropy $h({\bf Z})$ and entropy power $P({\bf Z})$.  Furthermore, let $\tilde{\bf Z} \equiv \sqrt{\eta}\,\tilde{{\bf X}} + \sqrt{1-\eta}\,\tilde{{\bf Y}}$. Three equivalent forms of the Entropy Power Inequality (EPI), see, e.g., \cite{Rioul}, are then:
\begin{eqnarray}
P({\bf Z}) &\ge& \eta P({\bf X}) + (1-\eta)P({\bf Y})
\label{EPI1}\\[.12in]
h({\bf Z}) &\ge& h(\tilde{\bf Z}) 
\label{EPI2}\\[.12in]
h({\bf Z}) &\ge& \eta h({\bf X}) + (1-\eta) h({\bf Y}).
\label{EPI3}
\end{eqnarray}

\subsection{The Entropy Photon-Number Inequality}
Let $\hat{\boldsymbol a} = [\begin{array}{cccc} \hat{a}_1 & \hat{a}_2 & \cdots & \hat{a}_n\end{array}]$ and $\hat{\boldsymbol b} = [\begin{array}{cccc} \hat{b}_1 & \hat{b}_2 & \cdots & \hat{b}_n\end{array}]$ be vectors of photon annihilation operators for a collection of 2$n$ different electromagnetic field modes of frequency $\omega$ \cite{quantumoptics}. The joint state of the modes associated with $\hat{\boldsymbol a}$ and $\hat{\boldsymbol b}$ is given by the product-state density operator $\hat{\rho}_{\boldsymbol a\boldsymbol b} = \hat{\rho}_{\boldsymbol a}\otimes \hat{\rho}_{\boldsymbol b}$, where $\hat{\rho}_{\boldsymbol a}$ and $\hat{\rho}_{\boldsymbol b}$ are the density operators associated with the $\hat{\boldsymbol a}$ and $\hat{\boldsymbol b}$ modes, respectively.  The von Neumann entropies of the $\hat{\boldsymbol a}$ and $\hat{\boldsymbol b}$ modes are $S(\hat{\rho}_{\boldsymbol a}) = -{\rm tr}[\hat{\rho}_{\boldsymbol a}\ln(\hat{\rho}_{\boldsymbol a})]$ and
$S(\hat{\rho}_{\boldsymbol b}) = -{\rm tr}[\hat{\rho}_{\boldsymbol b}\ln(\hat{\rho}_{\boldsymbol b})]$. 

The thermal state of a mode with annihilation operator $\hat{a}$ has two equivalent definitions:
\begin{equation}
\hat{\rho}_T = \int\!{\rm d}^2\alpha\, \frac{e^{-|\alpha|^2/N}}{\pi N}\,|\alpha\rangle \langle \alpha|,
\label{cohstateform}
\end{equation}
and 
\begin{equation}
\hat{\rho}_T = \sum_{i=0}^\infty \frac{N^i}{(N+1)^{i+1}}\,|i\rangle \langle i|,
\label{numberstateform}
\end{equation}
where $N = \langle \hat{a}^\dagger \hat{a}\rangle$ is the average photon number.  In Eq.~(\ref{cohstateform}), $|\alpha\rangle$ is the coherent state of amplitude $\alpha$, i.e., it satisfies $\hat{a}|\alpha\rangle = \alpha|\alpha\rangle$, for $\alpha$ a complex number. In Eq.~(\ref{numberstateform}), $|i\rangle$ is the $i$-photon state, i.e., it satisfies
$\hat{N}|i\rangle = i|i\rangle$, for $i = 0, 1, 2, \ldots$, with $\hat{N} \equiv \hat{a}^\dagger\hat{a}$ being the photon number operator.  Physically, Eq.~(\ref{cohstateform}) says that the thermal state is an isotropic Gaussian mixture of coherent states.  Equation~(\ref{numberstateform}), on the other hand, says that the thermal state is a Bose-Einstein mixture of number states.  From Eq.~(\ref{numberstateform}) we immediately have that $S(\hat{\rho}_T) = g(N)$, because the photon-number states are orthonormal.\footnote[5]{The coherent states, $\{|\alpha\rangle\}$, are \em not\/\rm\ orthonormal, but rather overcomplete.} 

The entropy photon-numbers of the density operators $\hat{\rho}_{\boldsymbol a}$ and $\hat{\rho}_{\boldsymbol b}$ are defined as follows:
\begin{equation}
N(\hat{\rho}_{\boldsymbol a})\equiv g^{-1}(S(\hat{\rho}_{\boldsymbol a})/n)\, \mbox{ and }\,
N(\hat{\rho}_{\boldsymbol b})\equiv g^{-1}(S(\hat{\rho}_{\boldsymbol b})/n).
\end{equation}
Thus, if
$\hat{\rho}_{\tilde{\boldsymbol a}} \equiv \bigotimes_{i=1}^n \hat{\rho}_{T_{a_i}}$ and
$\hat{\rho}_{\tilde{\boldsymbol b}} \equiv \bigotimes_{i=1}^n \hat{\rho}_{T_{b_i}}$,
where $\hat{\rho}_{T_{a_i}}$ is the thermal state of average photon  number $N(\hat{\rho}_{\boldsymbol a})$ for the $\hat{a}_i$ mode and $\hat{\rho}_{T_{b_i}}$ is the thermal state of average photon number $N(\hat{\rho}_{\boldsymbol b})$ for the $\hat{b}_i$ mode, then we have $S(\hat{\rho}_{\tilde{\boldsymbol a}}) = S(\hat{\rho}_{\boldsymbol a})$ and $S(\hat{\rho}_{\tilde{\boldsymbol b}}) = S(\hat{\rho}_{\boldsymbol b})$. We define a new vector of photon annihilation operators,  
$\hat{\boldsymbol c} = [\begin{array}{cccc} \hat{c}_1 & \hat{c}_2 & \cdots & \hat{c}_n\end{array}]$,
by the convex combination
\begin{equation}
\hat{\boldsymbol c} \equiv \sqrt{\eta}\,\hat{\boldsymbol a} + \sqrt{1-\eta}\,\hat{\boldsymbol b},
\quad\mbox{for $0\le \eta \le 1$,}
\label{beamsplitter_def}
\end{equation}
and use $\hat{\rho}_{\boldsymbol c}$ to denote its density operator.
This is equivalent to saying that $\hat{c}_i$ is the output of a lossless beam splitter whose inputs, $\hat{a}_i$ and $\hat{b}_i$, couple to that output with transmissivity $\eta$ and reflectivity $1-\eta$, respectively.  

We can now state two equivalent forms of our conjectured Entropy Photon-Number Inequality (EPnI) \cite{footnote4}:
\begin{eqnarray}
N(\hat{\rho}_{\boldsymbol c})&\ge& \eta N(\hat{\rho}_{\boldsymbol a})+ 
(1-\eta)N(\hat{\rho}_{\boldsymbol b})\label{EPnI1}\\[.12in]
S(\hat{\rho}_{\boldsymbol c}) &\ge & S(\hat{\rho}_{\tilde{\boldsymbol c}}),
\label{EPnI2}
\end{eqnarray}
where $\hat{\rho}_{\tilde{\boldsymbol c}} \equiv \bigotimes_{i=1}^n \hat{\rho}_{T_{c_i}}$ with $\hat{\rho}_{T_{c_i}}$ being the thermal state of average photon  number ${\eta}N(\hat{\rho}_{\boldsymbol a})+(1-\eta)N(\hat{\rho}_{\boldsymbol b})$ for $\hat{c}_i$. 

\section{Minimum Output Entropy Conjectures}
By analogy with the classical EPI, we might expect there to be a third equivalent form of the quantum EPnI, viz.,
\begin{equation}
S(\hat{\rho}_{\boldsymbol c}) \ge \eta S(\hat{\rho}_{\boldsymbol a}) + (1-\eta) S(\hat{\rho}_{\boldsymbol b}).
\label{EPnI3}
\end{equation}
It is easily shown that (\ref{EPnI1}) implies (\ref{EPnI3}) \cite{footnote5}, but we have not been able to prove the converse.  Indeed, we suspect that the converse might be false. More important than whether or not (\ref{EPnI3}) is equivalent to (\ref{EPnI1}) and (\ref{EPnI2}), is the role of the EPnI in proving classical information capacity results for bosonic channels.  In particular, the EPnI provides simple proofs of the following two minimum output entropy conjectures.  These conjectures are important because proving minimum output entropy conjecture~1 also proves the conjectured capacity of the thermal-noise channel \cite{thermal}, and proving minimum output entropy conjecture~2 also proves the conjectured capacity region of the bosonic broadcast channel \cite{broadcast}.   Furthermore, as we have shown above, proving minimum output entropy conjecture~2 also establishes the privacy capacity of the bosonic wiretap channel and the single-letter quantum capacity of the lossy bosonic channel.  

{\bf{Minimum Output Entropy Conjecture 1 ---}}
Let ${\boldsymbol a}$ and ${\boldsymbol b}$ be $n$-dimensional vectors of annihilation operators, with joint density operator $\hat{\rho}_{\boldsymbol a\boldsymbol b} = (|\psi\rangle_{\boldsymbol a}{}_{\boldsymbol a}\langle \psi|)\otimes \hat{\rho}_{\boldsymbol b}$, where $|\psi\rangle_{\boldsymbol a}$ is an arbitrary zero-mean-field pure state of the ${\boldsymbol a}$ modes and  $\hat{\rho}_{\boldsymbol b} = \bigotimes_{i=1}^n\hat{\rho}_{T_{b_i}}$ with $\hat{\rho}_{T_{b_i}}$ being the $\hat{b}_i$ mode's thermal state of average photon number $K$.  
Define a new vector of photon annihilation operators,  
$\hat{\boldsymbol c} = [\begin{array}{cccc} \hat{c}_1 & \hat{c}_2 & \cdots & \hat{c}_n\end{array}]$,
by the convex combination~\eqref{beamsplitter_def}
and use $\hat{\rho}_{\boldsymbol c}$ to denote its density operator and $S(\hat{\rho}_{\boldsymbol c})$ to denote its von Neumann entropy.  Then choosing $|\psi\rangle_{\boldsymbol a}$ to be the $n$-mode vacuum state minimizes $S(\hat{\rho}_{\boldsymbol c})$.  

{\bf{Minimum Output Entropy Conjecture 2 ---}}
Let ${\boldsymbol a}$ and ${\boldsymbol b}$ be $n$-dimensional vectors of annihilation operators with joint density operator $\hat{\rho}_{\boldsymbol a\boldsymbol b} = (|\psi\rangle_{\boldsymbol a}{}_{\boldsymbol a}\langle \psi|)\otimes \hat{\rho}_{\boldsymbol b}$, where $|\psi\rangle_{\boldsymbol a} = \bigotimes_{i=1}^n|0\rangle_{a_i}$ is the $n$-mode vacuum state and 
$\hat{\rho}_{\boldsymbol b}$ has von Neumann entropy $S(\hat{\rho}_{\boldsymbol b}) = ng(K)$ for some $K \ge 0$.    
Define a new vector of photon annihilation operators,  
$\hat{\boldsymbol c} = [\begin{array}{cccc} \hat{c}_1 & \hat{c}_2 & \cdots & \hat{c}_n\end{array}]$,
by the convex combination~\eqref{beamsplitter_def}
and use $\hat{\rho}_{\boldsymbol c}$ to denote its density operator and $S(\hat{\rho}_{\boldsymbol c})$ to denote its von Neumann entropy.  Then choosing $\hat{\rho}_{\boldsymbol b} = \bigotimes_{i=1}^n\hat{\rho}_{T_{b_i}}$ with $\hat{\rho}_{T_{b_i}}$ being the $\hat{b}_i$ mode's thermal state of average photon number $K$ minimizes  $S(\hat{\rho}_{\boldsymbol c})$.

To see that the EPnI encompasses both of the preceding minimum output entropy conjectures is our final task in this paper. We begin by using the premise of conjecture 1 in (\ref{EPnI1}).  Because the $\hat{\boldsymbol a}$ modes are in a pure state, we get $S(\hat{\rho}_{\boldsymbol a})= 0$ and hence
the EPnI tells us that
\begin{equation}
N(\hat{\rho}_{\boldsymbol c}) \ge  (1-\eta)N(\hat{\rho}_{\boldsymbol b}) = (1-\eta)K.
\end{equation}
Taking $g(\cdot)$ on both sides of this inequality, we get $S(\hat{\rho}_{\boldsymbol c})/n \ge g[(1-\eta)K]$. 
But, if $|\psi\rangle_{\boldsymbol a}$ is the $n$-mode vacuum state, we can easily show that 
$\hat{\rho}_{\boldsymbol c} = \bigotimes_{i=1}^n\hat{\rho}_{T_{c_i}}$,
with $\hat{\rho}_{T_{c_i}}$ being the $\hat{c}_i$ mode's thermal state of average photon number $(1-\eta)K$.  Thus, when $|\psi\rangle_{\boldsymbol a}$ is the $n$-mode vacuum state we get
$S(\hat{\rho}_{\boldsymbol c}) = ng[(1-\eta)K]$, which completes the proof.   

Next, we apply the premise of conjecture~2 in (\ref{EPnI1}).  Once again, the $\hat{\boldsymbol a}$ modes are in a pure state, so we get
\begin{equation}
N(\hat{\rho}_{\boldsymbol c}) \ge  (1-\eta)N(\hat{\rho}_{\boldsymbol b}) = (1-\eta)K,
\end{equation}
and hence $S(\hat{\rho}_{\boldsymbol c})/n \ge g[(1-\eta)K]$. But, taking $\hat{\rho}_{\boldsymbol b} = \bigotimes_{i=1}^n\hat{\rho}_{T_{b_i}}$, with $\hat{\rho}_{T_{b_i}}$ being the $\hat{b}_i$ mode's thermal state of average photon number $K$, satisfies the premise of minimum output entropy conjecture~2 and implies that $\hat{\rho}_{\boldsymbol c} = \bigotimes_{i=1}^n\hat{\rho}_{T_{c_i}}$, with $\hat{\rho}_{T_{c_i}}$ being the $\hat{c}_i$ mode's thermal state of average photon number $(1-\eta)K$.  In this case we have $S(\hat{\rho}_{\boldsymbol c}) = ng[(1-\eta)K]$, which completes the proof.  

\section{Conclusion}

We conjectured a quantum version of the classical entropy power inequality, which subsumes two minimum output entropy conjectures that prior work has shown to be sufficient to prove the capacity of the point-to-point thermal-noise lossy  bosonic channel, and the bosonic broadcast channel respectively \cite{thermal, broadcast}. Even though proving this more general inequality---the Entropy Photon-number Inequality (EPnI)---might seem harder than the two minimum output entropy conjectures, there is a possibility of drawing parallels from the proofs of the classical entropy power inequality \cite{Rioul}. In this paper, we have also shown that the EPnI also implies the proof of the privacy capacity of the bosonic wiretap channel.  Furthermore, using a result from  \cite{graeme}, we have that the degraded nature of the bosonic wiretap channel implies that its privacy capacity equals the single-letterquantum capacity of the lossy bosonic channel.  Moreover. both of these capacities are achieved by coherent-state encoding using an isotropic Gaussian prior. 

\section*{Acknowledgements}

This research was supported by the W. M. Keck Foundation Center for Extreme Quantum Information Theory.


\begin{thebibliography}{2}
\bibitem{footnote4} To show that (\ref{EPnI1}) implies (\ref{EPnI2}), assume (\ref{EPnI1}) is true:
\begin{eqnarray}
N(\hat{\rho}_{\boldsymbol c})&\ge& \eta N(\hat{\rho}_{\boldsymbol a})+ 
(1-\eta)N(\hat{\rho}_{\boldsymbol b})\\[.12in]
&=& \eta N(\hat{\rho}_{\tilde{\boldsymbol a}}) + 
(1-\eta) N(\hat{\rho}_{\tilde{\boldsymbol b}})
\label{deriv12}
\end{eqnarray}
Now, if $\hat{\rho}_{\tilde{\boldsymbol a}\tilde{\boldsymbol b}} = \hat{\rho}_{\tilde{\boldsymbol a}} \otimes 
\hat{\rho}_{\tilde{\boldsymbol b}}$ is the joint density operator of the $\hat{\boldsymbol a}$ and $\hat{\boldsymbol b}$ modes, we find that the state of the $\hat{\boldsymbol c}$ modes is 
$\hat{\rho}_{\tilde{\boldsymbol c}} \equiv \bigotimes_{i=1}^n 
\hat{\rho}_{T_{c_i}}$,
where $\hat{\rho}_{T_{c_i}}$ is a thermal state with average photon number given by
$N(\hat{\rho}_{\tilde{\boldsymbol c}}) = \eta N(\hat{\rho}_{\tilde{\boldsymbol a}}) + (1-\eta) N(\hat{\rho}_{\tilde{\boldsymbol b}})$,
so that $S(\hat{\rho}_{\tilde{\boldsymbol c}}) = ng[N(\hat{\rho}_{\tilde{\boldsymbol c}})]$.  
Thus, from (\ref{deriv12}) we get $N(\hat{\rho}_{\boldsymbol c})\ge 
N(\hat{\rho}_{\tilde{\boldsymbol c}}) =  g^{-1}(S(\hat{\rho}_{\tilde{\boldsymbol c}})/n)$. Taking $g(\cdot)$ of both sides of this inequality completes the proof.

To show that (\ref{EPnI2}) implies (\ref{EPnI1}), assume (\ref{EPnI2}) is true:
\begin{eqnarray}
N(\hat{\rho}_{\boldsymbol c})&=& g^{-1}(S(\hat{\rho}_{\boldsymbol c})/n) \nonumber \\[.12in] 
&\ge & g^{-1}(S(\hat{\rho}_{\tilde{\boldsymbol c}})/n)  = 
g^{-1}[g(\eta N(\hat{\rho}_{\tilde{\boldsymbol a}}) + (1-\eta) N(\hat{\rho}_{\tilde{\boldsymbol b}}))] \nonumber \\[.12in]
&=& \eta N(\hat{\rho}_{\tilde{\boldsymbol a}}) + (1-\eta) N(\hat{\rho}_{\tilde{\boldsymbol b}}) \nonumber \\[.12in]
&=& \eta N(\hat{\rho}_{\boldsymbol a})+ (1-\eta) N(\hat{\rho}_{\boldsymbol b}),
\end{eqnarray}
where the inequality is due to $g^{-1}(S)$ being a monotonically increasing function of $S$, and the proof is complete.
\bibitem{footnote5}Assume that  (\ref{EPnI1}) is true. We then have that $N(\hat{\rho}_{\boldsymbol c}) \ge \eta N(\hat{\rho}_{\boldsymbol a})+ (1-\eta)N(\hat{\rho}_{\boldsymbol b})$, so that
\begin{eqnarray}
S(\hat{\rho}_{\boldsymbol c}) &=& ng[N(\hat{\rho}_{\boldsymbol c})] \ge ng[\eta N(\hat{\rho}_{\boldsymbol a}) + (1-\eta) N(\hat{\rho}_{\boldsymbol b})] \\[.12in] 
&\ge& \eta ng[N(\hat{\rho}_{\boldsymbol a})] + (1-\eta)ng[N(\hat{\rho}_{\boldsymbol b})] \\[.12in]
&=& \eta S(\hat{\rho}_{\boldsymbol a}) + (1-\eta)S(\hat{\rho}_{\boldsymbol b}),
\end{eqnarray}
where the second inequality follows from $g(N)$ being concave, and the proof is complete.
\end{thebibliography}
\end{document}